\documentclass[aps,prb,twocolumn,floatfix,showpacs]{revtex4-1}
\usepackage{graphicx}
\usepackage{amsfonts}
\usepackage{amssymb}
\usepackage{amsmath}
\usepackage{enumerate}
\usepackage{multirow}
\usepackage{color}
\usepackage[all]{xy}
\usepackage{epstopdf}

\newcommand{\be}{\begin{equation}}
\newcommand{\ee}{\end{equation}}
\newcommand{\ba}{\begin{eqnarray}}
\newcommand{\ea}{\end{eqnarray}}
\newcommand{\baa}{\begin{eqnarray*}}
\newcommand{\eaa}{\end{eqnarray*}}

\usepackage{amssymb}
\usepackage{graphicx}
\usepackage{amsmath}
\usepackage{dsfont}
\usepackage{bbm}
\begin{document}
\title{Majorana fermions in honeycomb lattices}
\author{C. Dutreix}

\author{M. Guigou}

\author{D. Chevallier}

\author{C. Bena}

\affiliation{Laboratoire de Physique des Solides, Orsay, France}
\affiliation{Institut de Physique Th\'eorique, CEA/Saclay, Orme des Merisiers, 91190 Gif-sur-Yvette Cedex, France}

\begin{abstract}
We study the formation of Majorana fermions in honeycomb-lattice structures in the presence of a Zeeman field, Rashba spin-orbit coupling, and in the proximity of an s-wave superconductor. We show that an exact mapping exists between an anisotropic hexagonal-lattice nanoribbon at $k=0$ and a one-dimensional chain, for which the existence of Majorana fermions has been extensively discussed. Consequently we can predict the conditions for the emergence of Majorana fermions at the edges of such ribbon, and relate the existence of Majoranas to a band inversion in the bulk band structure. Moreover we find that similar situations arise in isotropic lattices and we give some examples which show the formation of Majorana fermions in these structures.

\end{abstract}


\maketitle

\section{Introduction}
 
Majorana fermionic states, first introduced in 1937 by E. Majorana \cite{ettore} as real solutions of the Dirac equation, have recently generated a lot of interest in condensed matter physics because of their exotic properties \cite{beenakker, alicea}. Unlike the usual Dirac particles, the Majorana fermions obey non-Abelian statistics, which open the perspective of their use in topological quantum computation \cite{nayak_dassarma}. Their existence has been predicted in various systems: fractional quantum Hall systems \cite{Moore91,Read00}, topological insulators \cite{Fu08,Tanaka09}, optical lattices\cite{Sato09}, p-wave superconductors \cite{Potter11,Wong12}, insulating \cite{Cook11} and semiconducting nanowires with strong spin-orbit coupling such as InAs or InSb \cite{lutchyn_dassarma,oreg_vonoppen,mourik12}. In particular, the latter are expected to exhibit end Majorana fermionic states in the presence of an applied Zeeman field and in the proximity of an s-wave superconductor (SC). A topological nanowire can be modeled as a one-dimensional (1D) tight-binding chain, for which Majorana fermions can form in certain regimes of parameters \cite{kitaev}. Besides, several studies of various two-dimensional (2D) systems have pointed out the possibility to generate Majorana fermions inside a vortex core as a consequence of symmetry breaking, e.g. due to the presence of an impurity, as well as diffusive Majorana edge states \cite{2D_diffusive,mudry,Sato11,Wong12}. However, most of these studies rely on square-lattice models. The study of Majorana fermions in a honeycomb lattice, like that of graphene, has received less attention \cite{mudry,blackschaffer,klinovaja12,klinovaja13a,klinovaja13b}.

The most known and studied material exhibiting a honeycomb-lattice structure is graphene; this is one of the most intriguing materials discovered during the last decade \cite{novoselov_geim}. Its elementary excitations are Dirac fermions, massless relativistic particles described by a 2D Dirac Hamiltonian. The exotic properties of graphene  led to an extensive theoretical and experimental study focusing on both monolayer and bilayer graphene, and taking into account features such as straining and curvature, disorder, etc. (see Ref. \onlinecite{honeycomb lattice_general_work} and references therein).


The possibility to get Majorana fermionic states in honeycomb-lattice structures is the main topic of this paper. In particular, we focus on a nanoribbon (NR) with zigzag edges. The properties of such ribbon can be described by mapping it to a one-dimensional dimer chain model. Taking inspiration from the analysis of a 1D wire \cite{kitaev}, for which the Majorana physics is well understood, we study the possibility of emergence of Majorana fermions when the NR is brought in the vicinity of an s-wave superconducting substrate, in the presence of Rashba spin-orbit coupling and of an in-plane Zeeman field. In particular, at $k=0$, and for a specific value of anisotropy corresponding to the merging of the two Dirac cones \cite{hasegawa06,montambaux09}, the dimer chain is exactly mapped onto a simple one-dimensional monomer chain describing a topological nanowire, for which the Majorana physics has been extensively studied \cite{lutchyn_dassarma,oreg_vonoppen}. Thus, in this limit, we can predict exactly the parameter regimes that support the existence of Majorana fermions in the anisotropic NR. 

We show that Majorana fermionic edge states can form in various other situations, for both anisotropic and isotropic lattices, and we provide some examples of such situations. We test that these zero-energy states are indeed Majoranas by using a tight-binding model and an exact diagonalization numerical technique which allows us to evaluate their wave-functions and calculate the Majorana polarization introduced in Refs.~\onlinecite{sticlet12,chevallier12}.

The paper is organized as follows. 
In section II we review the tight-binding model of the honeycomb lattice, we introduce the Hamiltonians of the extra physical processes that may yield to the formation of Majorana fermions, the Rashba spin-orbit coupling, the Zeeman field, and the s-wave superconductivity, and we describe the corresponding band structure for an infinite system.
In section III we consider finite size nanoribbons and we study the conditions for the emergence of Majorana edge states in these systems. Section III A describes the mapping of a two-dimensional zigzag NR onto a one-dimensional dimer chain. Section III B is devoted to the particular value of anisotropy for which the dimer chain becomes a simple one-dimensional monomer chain, and the system can be solved exactly. In Section III C we provide examples of forming Majorana fermions in an isotropic NR with zigzag edges. We conclude in section IV.

\section{Infinite system}

\subsection{The honeycomb lattice}
\begin{figure}[!h]
\includegraphics[width=7cm]{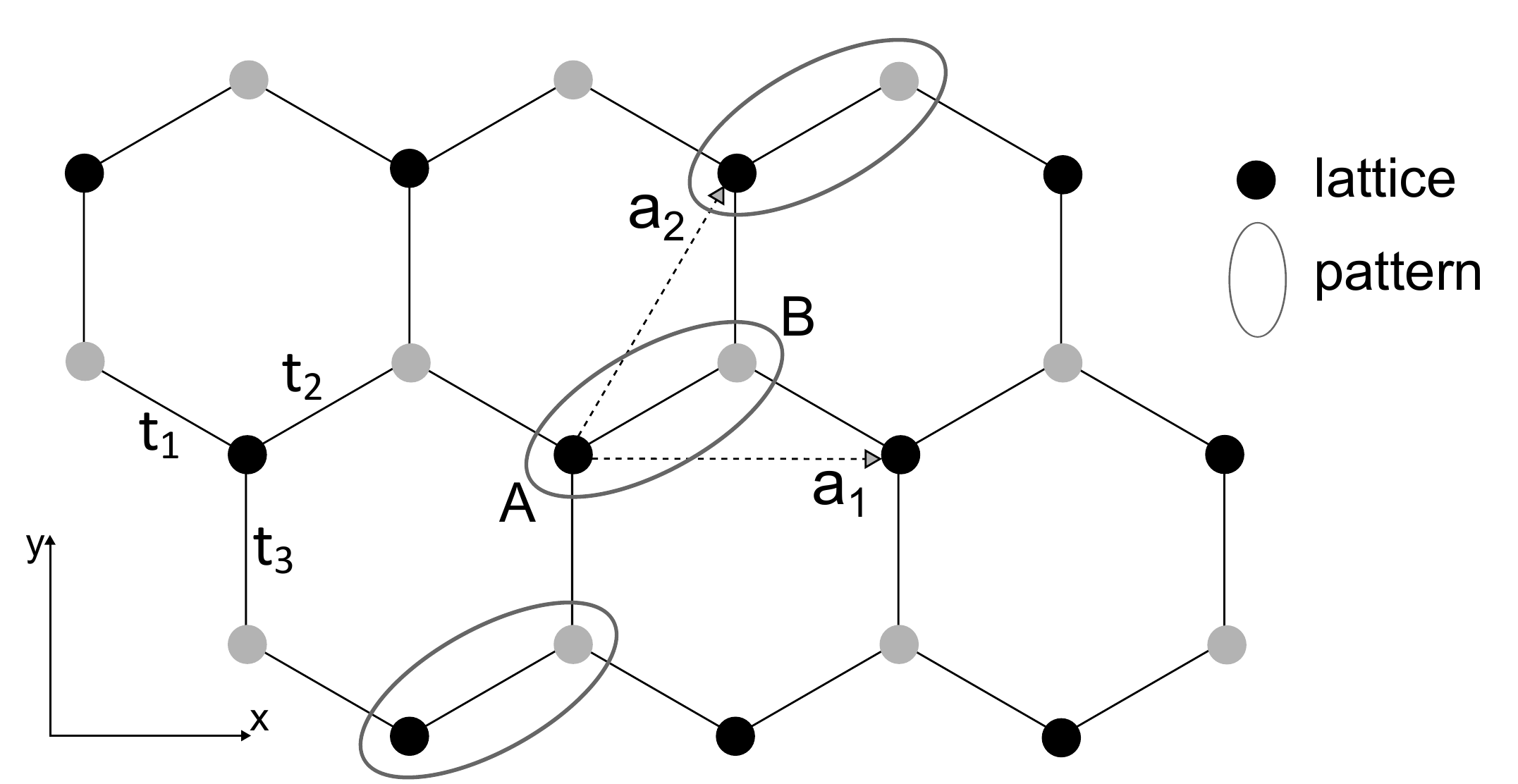}
\caption{\small Honeycomb lattice.}
\label{HoneycombLattice}
\end{figure}

We first review the band structure of the honeycomb lattice. Let us denote ${\bf{a_1}}=(\sqrt{3};0)a_0$ and ${\bf{a_2}}=(\sqrt{3};3)a_0/2$ the basis vectors that span the triangular Bravais lattice. The interatomic distance $a_0$, which for graphene is  0.142nm, is taken to be $a_0=1$ in what follows. In this basis, each unit cell is described by a pair of integers $(m,n)$, or equivalently by ${\bf{r}}_{i}=m{\bf{a_1}}+n{\bf{a_2}}$, and contains two inequivalent atoms, labeled A and B, as depicted in Fig. \ref{HoneycombLattice}. 

The bulk band structure of a honeycomb lattice is well described using a tight-binding approach in which each electron may hop between two sites $i$ and $j$ with a given amplitude $t_{ij}$. In this work only nearest neighbor hopping processes are considered, with fixed amplitudes $t_1=t_2=t$ and a variable amplitude $t_3$. The corresponding second-quantized Hamiltonian in real space is given by
\begin{align}
{\cal{H}}_{0}&=\sum_{\langle i,j\rangle,\sigma} t_{ij}a^{\dagger}_{i\sigma}b_{j\sigma} +h.c. \notag \\
		  &+ \mu \sum_{i,\sigma} a^{\dagger}_{i\sigma}a_{i\sigma}+b^{\dagger}_{i\sigma}b_{i\sigma},   \notag \\		   
\label{Hopping Hamiltonian}
\end{align}
where $\langle ., .\rangle$ denotes the nearest neighbor atoms, $\mu$ is the chemical potential, and $a_{i\sigma}$ (resp. $b_{i\sigma}$) annihilates an electron with spin $\sigma$ on an atom A (resp. B) in the unit cell defined by ${\bf{r}}_{i}$. 

\begin{figure}[t]
\centering
$\begin{array}{cc}
   \includegraphics[trim = 7mm 5mm 25mm 20mm, clip, width=4.3cm]
{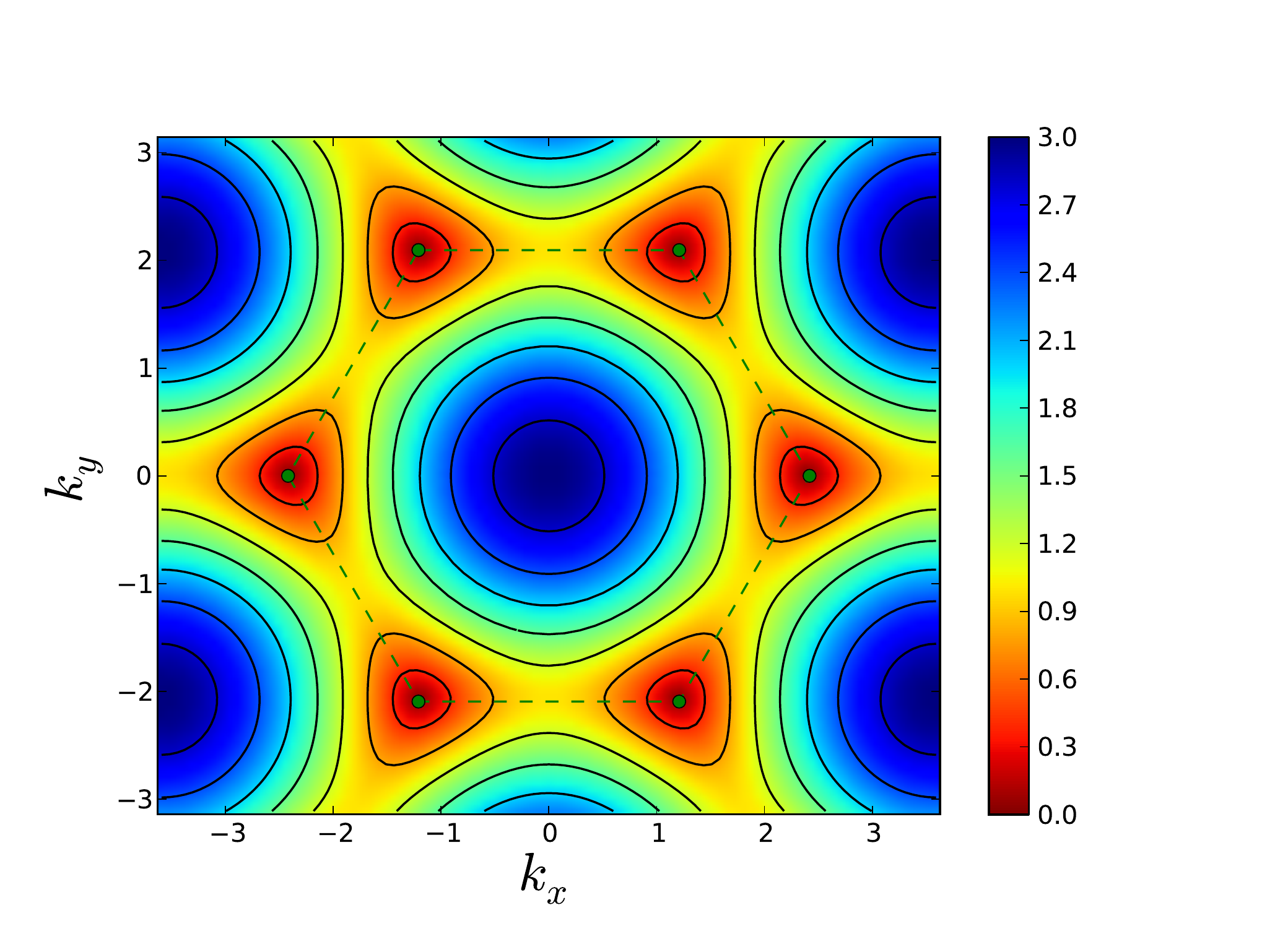}
   \includegraphics[trim = 7mm 5mm 25mm 20mm, clip, width=4.3cm]{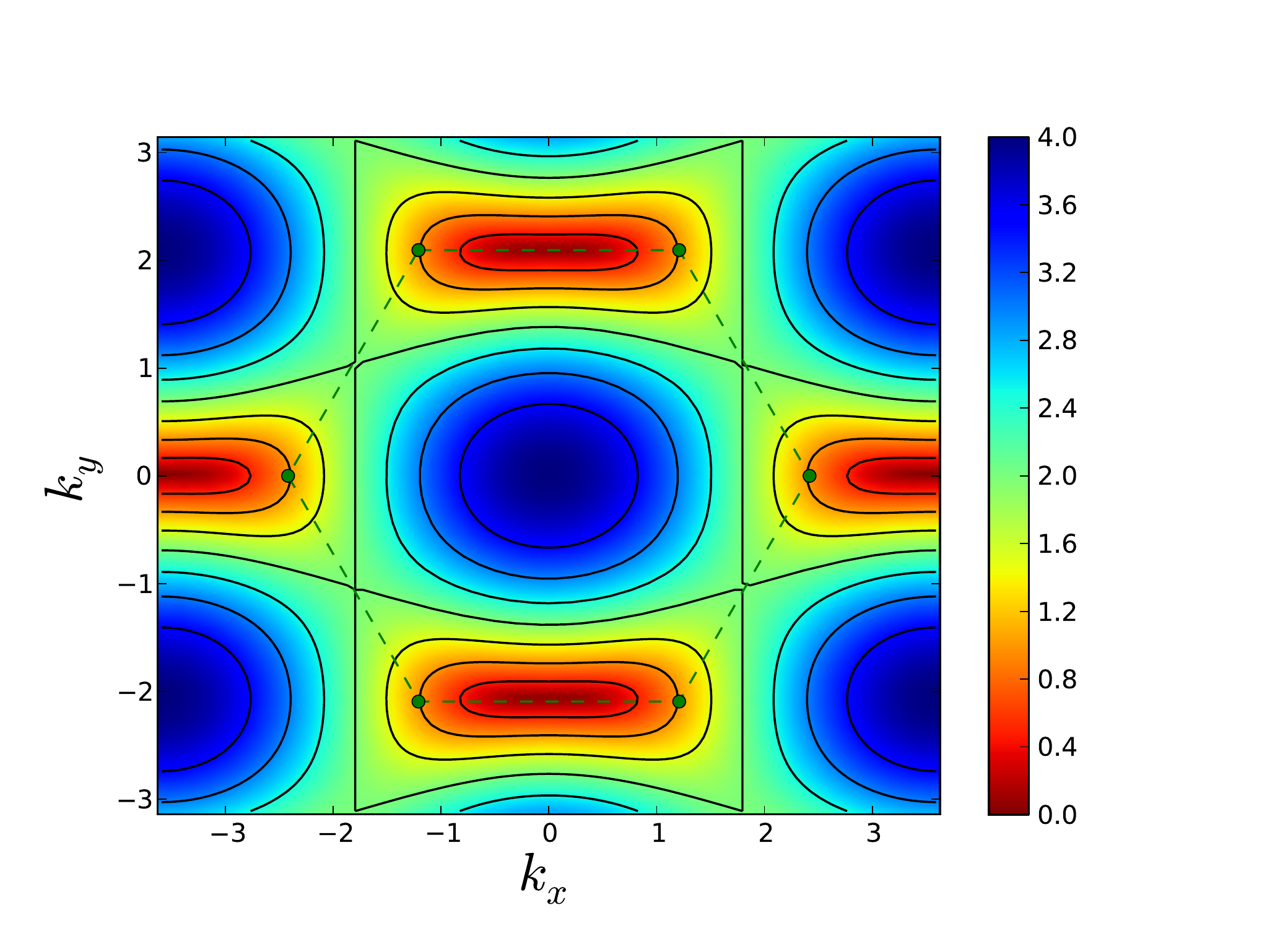}\\
\end{array}$
\caption{(Color online) Energy spectrum of a honeycomb lattice (in units of $t$) for $\mu=0$ when $t_3=t$ (left) and $t_3=2t$ (right). The latter corresponds to the merging of the Dirac points into a single time-reversal invariant point. The green dashed line depicts the Brillouin zone.}
\label{Bulk Spectra}
\end{figure}

The corresponding energy spectrum is presented in Fig. \ref{Bulk Spectra} for two particular values of the hopping amplitude $t_3$. When $t_3=t$, two inequivalent Fermi points lie at the corners of the Brillouin zone, and the dispersion relation around these points is conical (the red areas in the figure). For the peculiar value $t_3=2t$, these Dirac fermions merge into a single time-reversal invariant point where the dispersion remains linear along the y-axis, whereas it becomes quadratic along the x-axis\cite{hasegawa06,montambaux09}. Although these fermions are responsible for the peculiar properties, such as edge states and a different type of quantum Hall effect, in what follows we focus mainly on the dispersion around the zero momentum (the blue region in Fig. \ref{Bulk Spectra}), as it is this region of momentum space which controls the formation of Majorana fermions.

\subsection{S-wave pairing, Rashba spin-orbit interaction and Zeeman field}
As shown in previous works \cite{Fu08,Tanaka09,Sato09,Potter11,Wong12,Cook11,lutchyn_dassarma,oreg_vonoppen,kitaev,sticlet12,chevallier12}, Majorana fermions appear at the boundaries of topological superconductors. In order to study their emergence in honeycomb-lattice structures, we consider processes similar to those proposed to form Majoranas in a one-dimensional monomer chain \cite{alicea,lutchyn_dassarma,oreg_vonoppen}, i.e. applying a Zeeman field to a system with strong spin-orbit interaction and in the vicinity of an s-wave superconductor.


In the most well-known hexagonal-lattice system, graphene, while the Rashba spin-orbit interaction (SOI) is naturally present, its strength is very weak \cite{Huertas06,min06,yao07}. That can be enhanced by different methods, \textit{e.g.} applying a perpendicular electric field \cite{Efetov10,Zarea09} or a non-uniform magnetic field \cite{klinovaja13a}, tuning the local curvature of the sheet \cite{Huertas06}, or doping by 3d or 5d transition metal adatoms \cite{McChesney10,Hu12,Shestov12}. The Rashba SOI Hamiltonian can be written as
\begin{align}
{\cal{H}}_{R}&= i\lambda \sum_{\langle i,j\rangle,\sigma,\sigma'}  ({\bf d}_{ij} \times \boldsymbol{\sigma})_{\sigma,\sigma'}.{\bf e_z} a^{\dagger}_{i\sigma}b_{j\sigma'} +h.c., 
\label{Rashba Hamiltonian}
\end{align}
where the Rashba spin orbit-coupling of strength $\lambda$ tends to align the spins in a direction defined by the nearest neighbor vectors ${\bf d}_{ij}$. The vector $\boldsymbol{\sigma}$ describes the spin Pauli matrices and ${\bf e_z}$ is an out-of-plane unit vector. 

The s-wave superconductivity is induced in the system by proximity effect and is described by a uniform on-site superconducting order parameter $\Delta$ whose Hamiltonian can be written as
\begin{align}
{\cal{H}}_{S}&= \Delta \sum_{i,\sigma}  a^{\dagger}_{i,\sigma}a^{\dagger}_{i,-\sigma} + b^{\dagger}_{i,\sigma} b^{\dagger}_{i,-\sigma} +h.c..
\label{Superconducting Hamiltonian}
\end{align}

Finally, a magnetic field $\boldsymbol{B}$ is introduced, resulting in an out-of-plane Zeeman potential $V_Z=g\mu_B|\boldsymbol{B}| /2$, with $g$ the Land\'e g-factor and $\mu_B$ the Bohr magneton. This Zeeman Hamiltonian is described as follows
\begin{align}
{\cal{H}}_{Z}&= V_Z \sum_{i,\sigma,\sigma'}  a^{\dagger}_{i,\sigma}\sigma^{z}_{\sigma, \sigma'} a_{i,\sigma'} + b^{\dagger}_{i,\sigma}\sigma^{z}_{\sigma, \sigma'} b_{i,\sigma'}.
\label{Zeeman Hamiltonian}
\end{align}

\subsection{Band Structure}
\begin{figure}[t]
\centering
$\begin{array}{c}
\includegraphics[width=7cm]{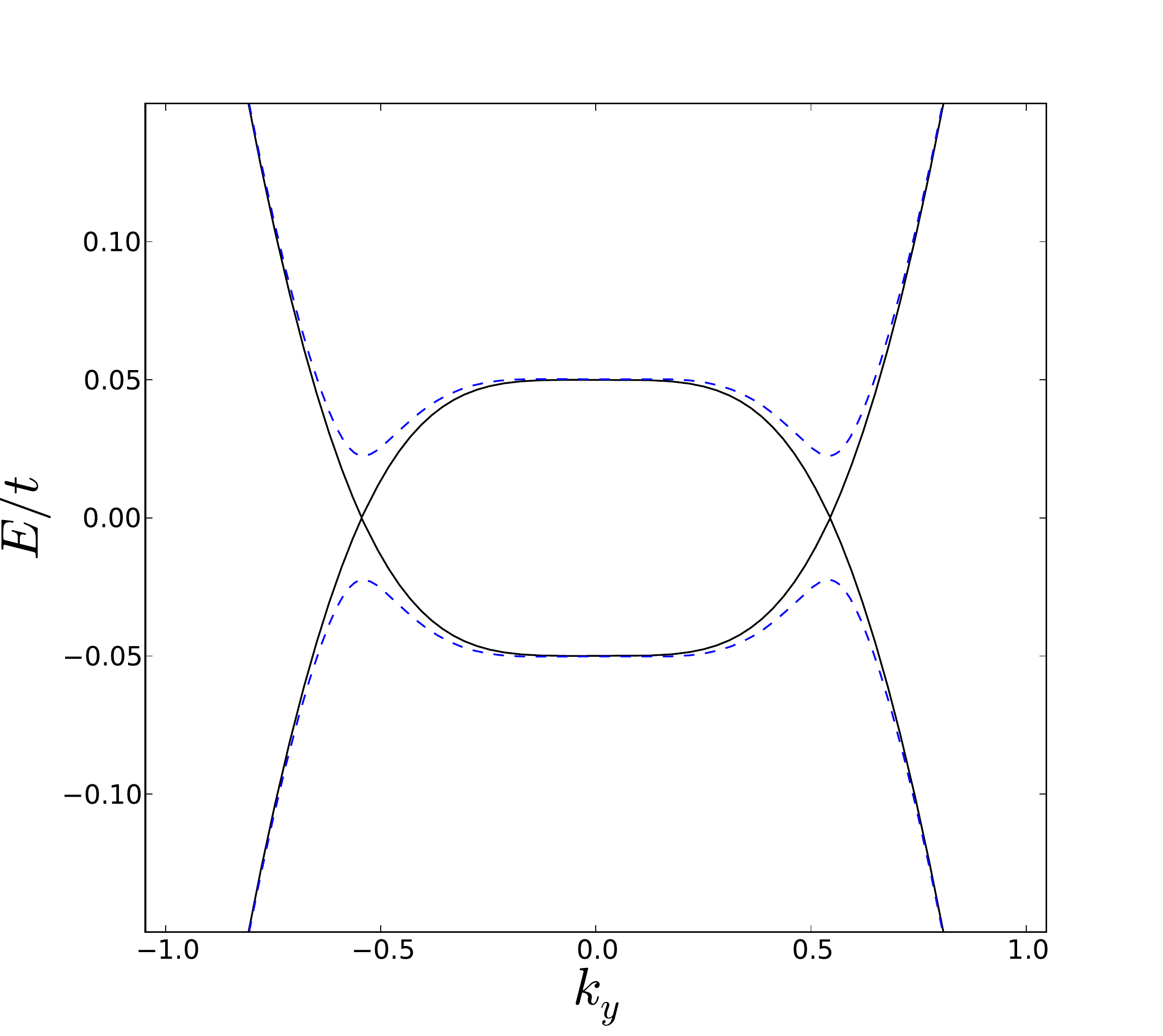}\\
\includegraphics[width=7cm]{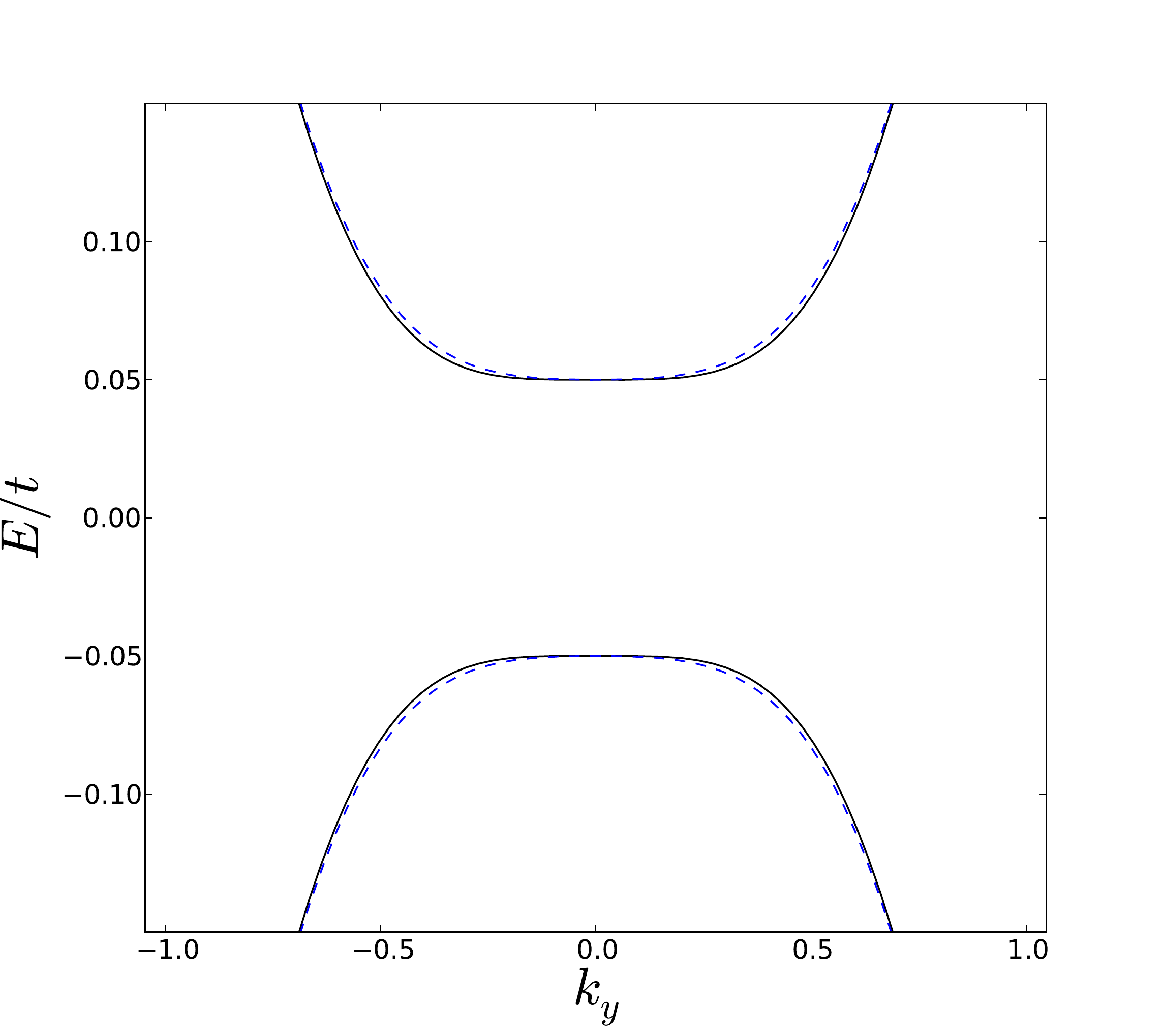}\\
\end{array}$
\caption{\small (Color online) Excitation spectrum as a function of $k_{y}$ when $k_{x}=0$ and $\mu=3t$, (upper panel) in the non trivial topological regime ($\Delta=0.45t$ and $V_Z=0.5t$), and (lower panel) in the trivial regime ($\Delta=0.55t$ and $V_Z=0.5t$) for $\lambda=0$ (black solid line) and for $\lambda=0.1t$ (blue dashed line).}
\label{ky-Spectrum}
\end{figure}

The electronic structure of an infinite honeycomb lattice in the presence of these processes can be a good starting point to predict in what circumstances Majorana fermions may form in a nanoribbon. The bulk band structure of the complete system is presented in Fig. \ref{ky-Spectrum} for two sets of parameter values. One important ingredient is the chemical potential which we take to be of the order of the bandwidth, $\mu=\sum_{i=1,2,3} t_i$. This choice will be justified rigorously in the next section. Here we only note that, under these circumstances, each energy band is ellipsoidal in the vicinity of ${\bf k}=0$, such that the gap closing occurs at this zero momentum in the spectrum and is given by the condition
\begin{align}\label{Gap closing condition}
\Big(\mu-\sum_{i=1,2,3}t_i\Big)^{2}+\Delta^{2}= V_Z^{2}
\end{align}
For instance, the dispersion relation depicted in Fig. \ref{ky-Spectrum} exhibits zero-energy band crossings if $\Delta < V_Z$. For such values of parameters, one expects that the introduction of edges in the system will give rise to new states inside the gap that do not correspond to any of the bands in the bulk spectrum. On the contrary, the dispersion relation plotted in  Fig. \ref{ky-Spectrum} for $\Delta>V_Z$ exhibits no band inversion at zero energy, and consequently no extra states besides the bulk states are expected to show up by the introduction of edges. We conjecture that this is a general feature of this model if the Rashba SOI is sufficiently small not to affect the band inversion mechanism. As we show in what follows, these new states are actually Majorana edge states. 
Thus, the infinite-system band structure provides us with a good starting point to search for Majorana fermions in a finite size nanoribbon. 


\section{Finite system: Formation of Majorana edge states}
\subsection{Description of a honeycomb lattice ribbon as a one-dimensional dimer chain}

\begin{figure}[!h]
\includegraphics[width=7cm]{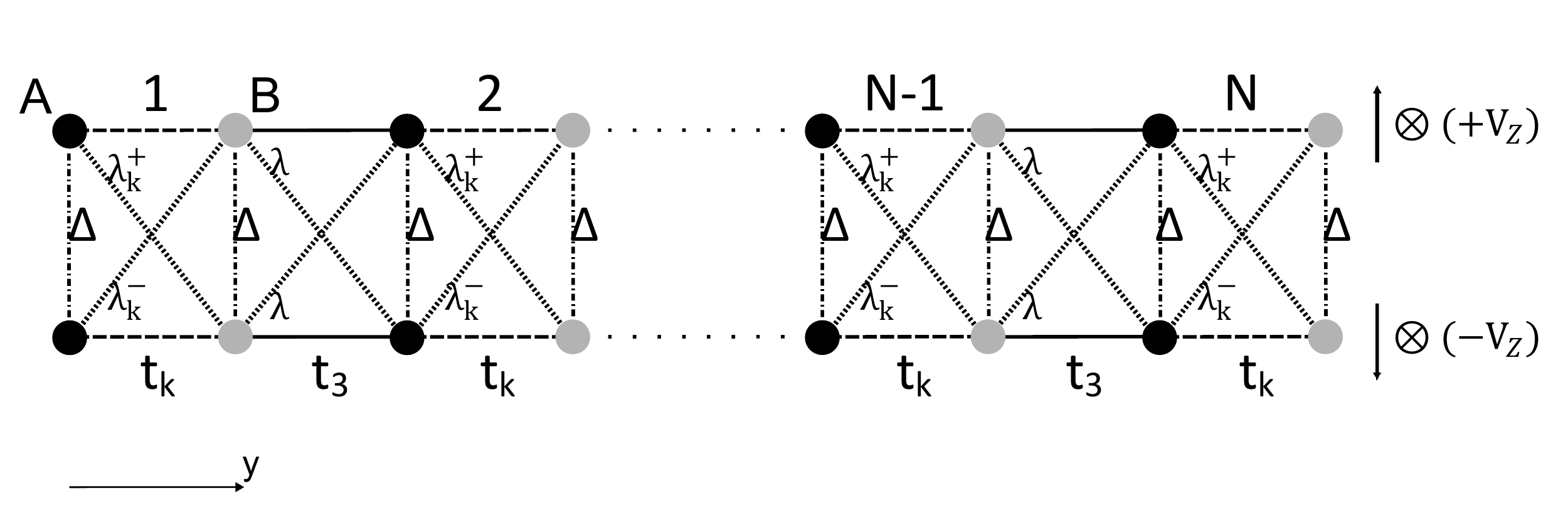}
\caption{\small 1D dimer chain: the upper (respectively lower) sub-chain is composed by N dimers whose spins point in the direction of $V_Z$ (respectively in the opposite direction). The direct connection between two identical atoms with opposite spin is given by the on-site pairing potential $\Delta$. The parameter $t_k$ represents the k-dependent hopping amplitude inside a dimer, while $t_3$ labels the hopping amplitude between dimers. Finally, the Rashba SOI connects different atoms of the two sub-chains, the corresponding amplitudes being $\lambda,\lambda_k^{\pm}$.}
\label{Mapping1D}
\end{figure}

We now introduce edges to the honeycomb-lattice structure and describe the resulting ribbon as a 1D chain of $N$ dimers. The system is constructed as follows: iterating periodically the AB dimer pattern (presented in Fig. \ref{HoneycombLattice}) $N$ times along the $\bf{a_2}$-direction leads to a chain of $N$ dimers, which is then infinitely repeated along the $\bf{a_1}$-direction. The ribbon has two zigzag edges, one made of A atoms only, and the opposite made of B atoms exclusively. Assuming a tight-binding model with only nearest neighbor hopping, and with an intra-dimer hopping parameter which depends on the momentum component parallel to the edges, the second quantized total Hamiltonian can be written as 

\begin{align}
{\cal{H}}_{TOT}&=\int_{-\pi}^{\pi} \frac{dk}{2\pi} \sum_{\sigma,n=1}^{N}\Big\{\mu\big[a^{\dagger}_{\sigma}(k,n)a_{\sigma}(k,n)+b^{\dagger}_{\sigma}(k,n)b_{\sigma}(k,n)\big]\notag \\
		     &+ t_ka^{\dagger}_{\sigma}(k,n)b_{\sigma}(k,n) +t_3a^{\dagger}_{\sigma}(k,n)b_{\sigma}(k,n-1)\notag \\
		     &+ i\big[\lambda^{\sigma}_{k} a^{\dagger}_{\sigma}(k,n)b_{-\sigma}(k,n)+\lambda^{\sigma}       a^{\dagger}_{\sigma}(k,n)b_{-\sigma}(k,n-1)\big] \notag \\
		      &+ {\rm sign}(\sigma) V_Z\big[ a^{\dagger}_{\sigma}(k,n)a_{\sigma}(k,n)+ b^{\dagger}_{\sigma}(k,n)b_{\sigma}(k,n)\big] \notag\\
	  &+ \Delta \big[a^{\dagger}_{\sigma}(k,n)a^{\dagger}_{-\sigma}(k,n) + b^{\dagger}_{\sigma}(k,n)b^{\dagger}_{-\sigma}(k,n)\big] +h.c.\Big\},
\label{1DtotalHamiltonian}
\end{align}
with $t_k=t_1e^{ik}+t_2$, and $k=\frac{k_x a_0}{\sqrt3}$ the dimensionless quantum number related to the translational invariance along the x-direction. Also $\lambda^{\sigma}_{k}=\lambda [1+e^{ik}+{\rm sign}(\sigma)i\sqrt3(1-e^{ik})]a_{0}/2$, and $\lambda^{\sigma}=-\lambda a_{0}$. Thus, the initial 2D system is described as a 1D chain of N dimers, for each value of $k$ (see Fig. \ref{Mapping1D}).

%
%
%

\subsection{Majorana modes in anisotropic honeycomb-lattice structures}

As it can be seen from Eq.~(\ref{1DtotalHamiltonian}) and Fig. \ref{Mapping1D}, at zero momentum ($k_x=k_y=0$) the system is equivalent to a 1D monomer chain when $t_3=2t$. At this particular point the two-dimensional anisotropic-honeycomb-lattice Majorana problem is thus equivalent to the one addressed in the literature for a one-dimensional chain \cite{lutchyn_dassarma,oreg_vonoppen}. For the one-dimensional finite-size system at $\mu=0 t$ (which corresponds to $\mu=4t$ for the NR) it has been shown that a zero-energy pair of end Majorana states forms for $\Delta<V_Z$ \cite{lutchyn_dassarma,oreg_vonoppen,sticlet12,chevallier12}. Therefore, the existence of Majorana states in the two-dimensional anisotropic nanoribbon should be controlled by the same criterion.

\begin{figure}[b]
\centering
$\begin{array}{c}
\includegraphics[width=7cm]{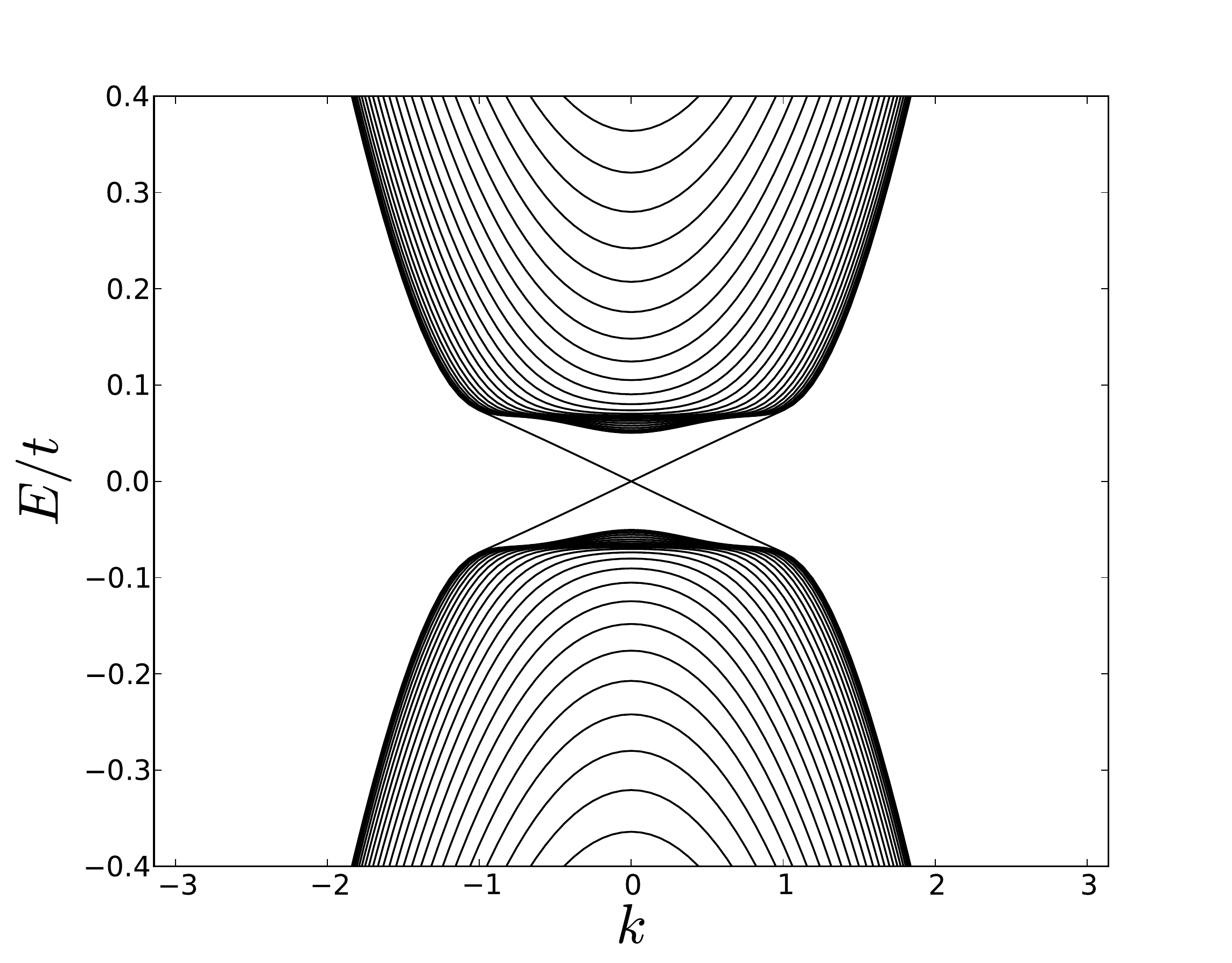}\\
\includegraphics[width=7cm]{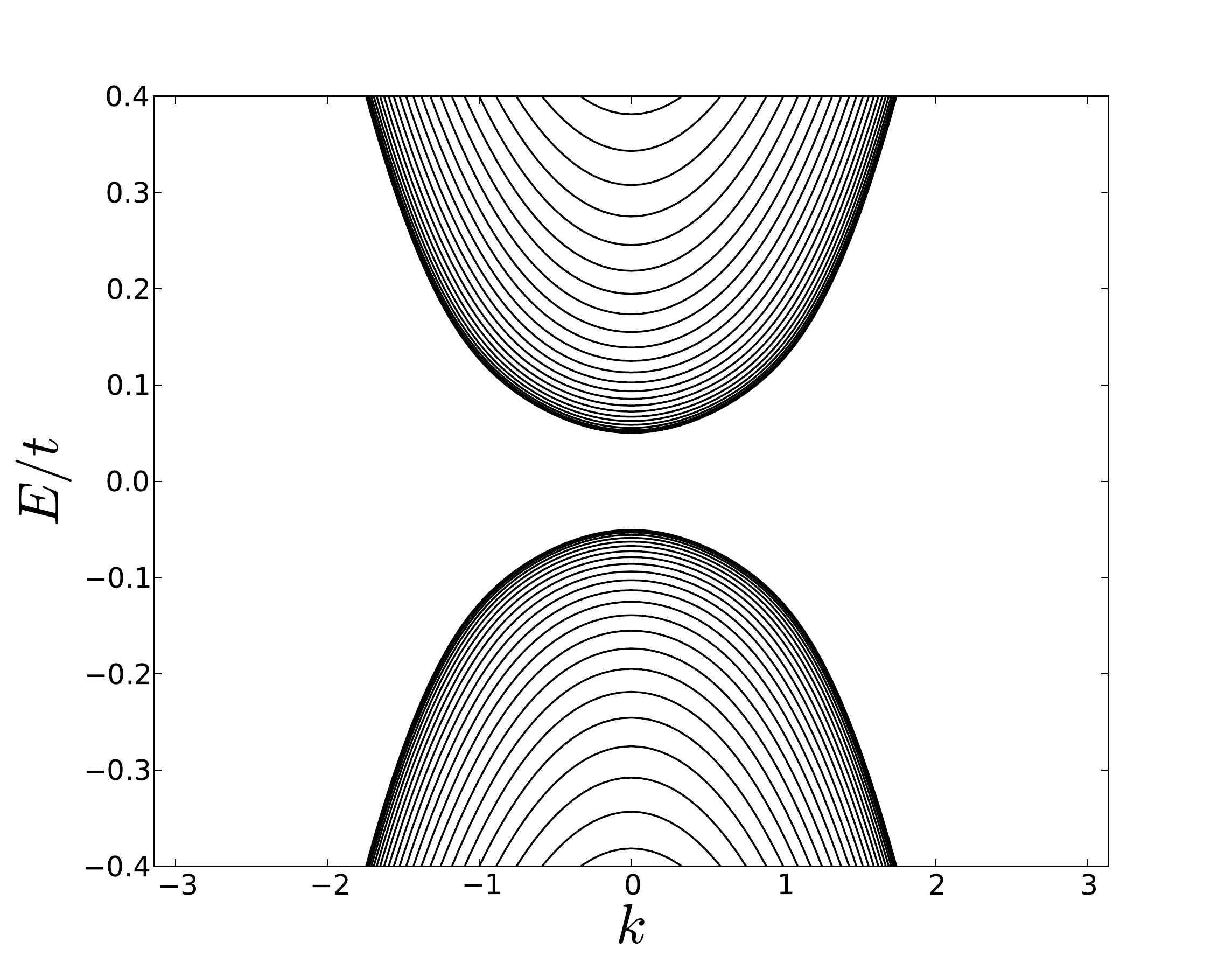}\\
\end{array}$
\caption{\small Band structure of an anisotropic zigzag NR with $N=50$  as a function of $k_{x}$. In the upper panel, the band structure in the topological phase ($\Delta=0.45t<V_Z=0.50t$), and in the lower panel the band structure in the trivial phase ($\Delta=0.55t>V_Z=0.50t$).}
\label{Band Structures}
\end{figure}

The NR band structures corresponding to both the topological and trivial phases ($\Delta<V_Z$ and respectively $\Delta>V_Z$) are plotted in Fig.~\ref{Band Structures}. As expected, two dispersive edge states crossing at zero energy appear if the system is in the topologically non-trivial phase, but not in the trivial phase. It is important to note that this criterion does not depend on the value of the Rashba coupling. Indeed, $\lambda^{\sigma}_k$ and $\lambda^{\sigma}$, introduced in the previous section, are equal at $k=0$, and similarly to the 1D chain, do not contribute quantitatively to the criterion of the existence of Majoranas.

%

A supplementary check to prove the Majorana character of the dispersive edge states appearing inside the gap is to study their wavefunctions and plot their Majorana polarizations.  Consistent with the definition introduced in Ref.~\onlinecite{sticlet12}, for a given eigenstate $\Phi^{\dagger}=(u_{A,\uparrow},u_{B,\uparrow},v_{A,\downarrow},v_{B,\downarrow},v_{A,\uparrow},v_{B,\uparrow},u_{A,\downarrow},u_{B,\downarrow})$, with $u$ and $v$ respectively the electron and hole amplitudes, the Majorana polarizations along the x- and y- axis are defined as
\begin{align}
P_{M_x} &= 2\textrm{Re}(u_{A,\downarrow}v^{*}_{A,\downarrow}-u_{A,\uparrow}v^{*}_{A,\uparrow}),\\
P_{M_y} &= 2\textrm{Im}(u_{A,\downarrow}v^{*}_{A,\downarrow}-u_{A,\uparrow}v^{*}_{A,\uparrow}).
\label{Polarizations}
\end{align}

\begin{figure}[b]
\centering
$\begin{array}{cc}
\includegraphics[trim = 100mm 0mm 20mm 0mm, clip, width=4cm]{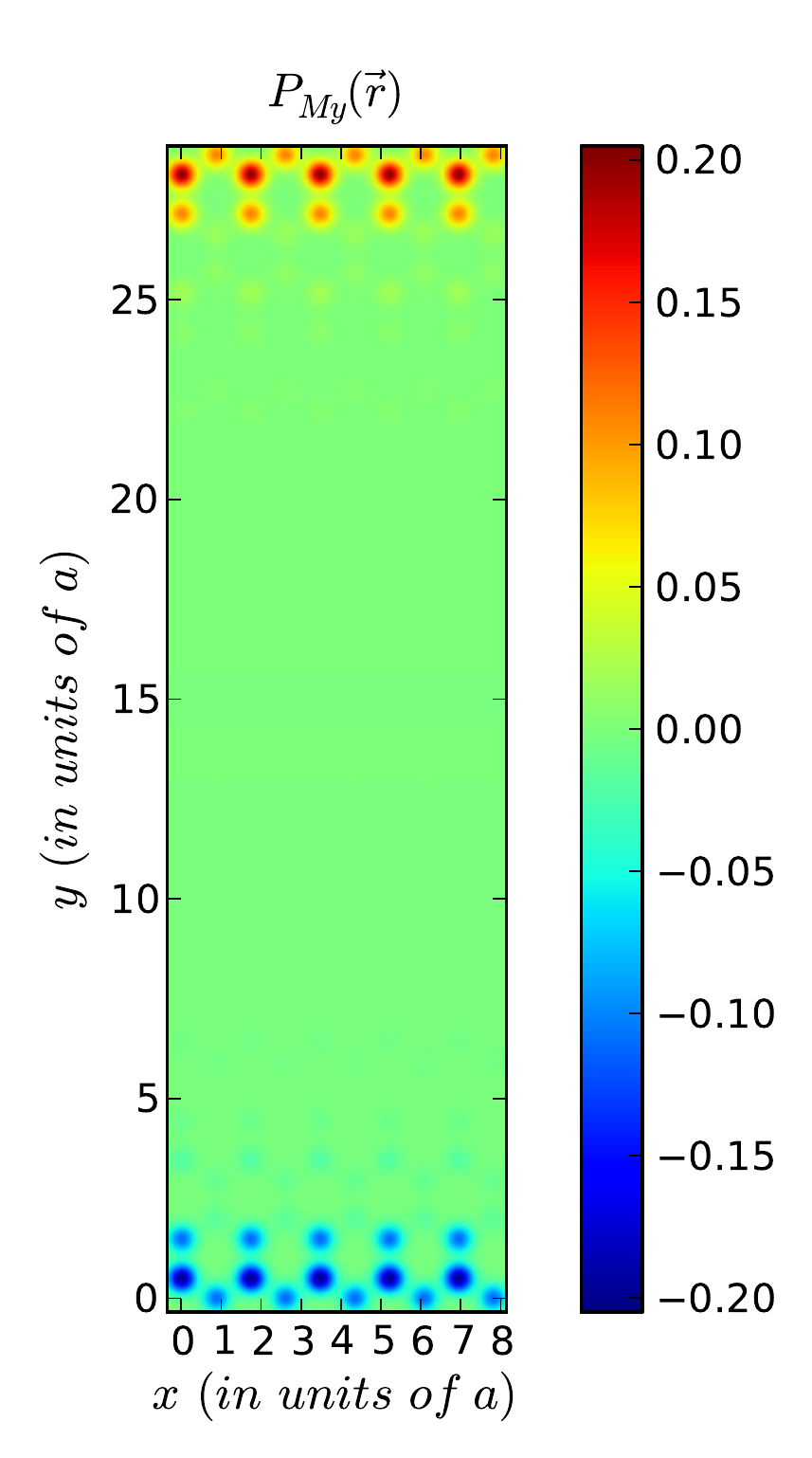}
\includegraphics[trim = 100mm 0mm 20mm 0mm, clip, width=4cm]{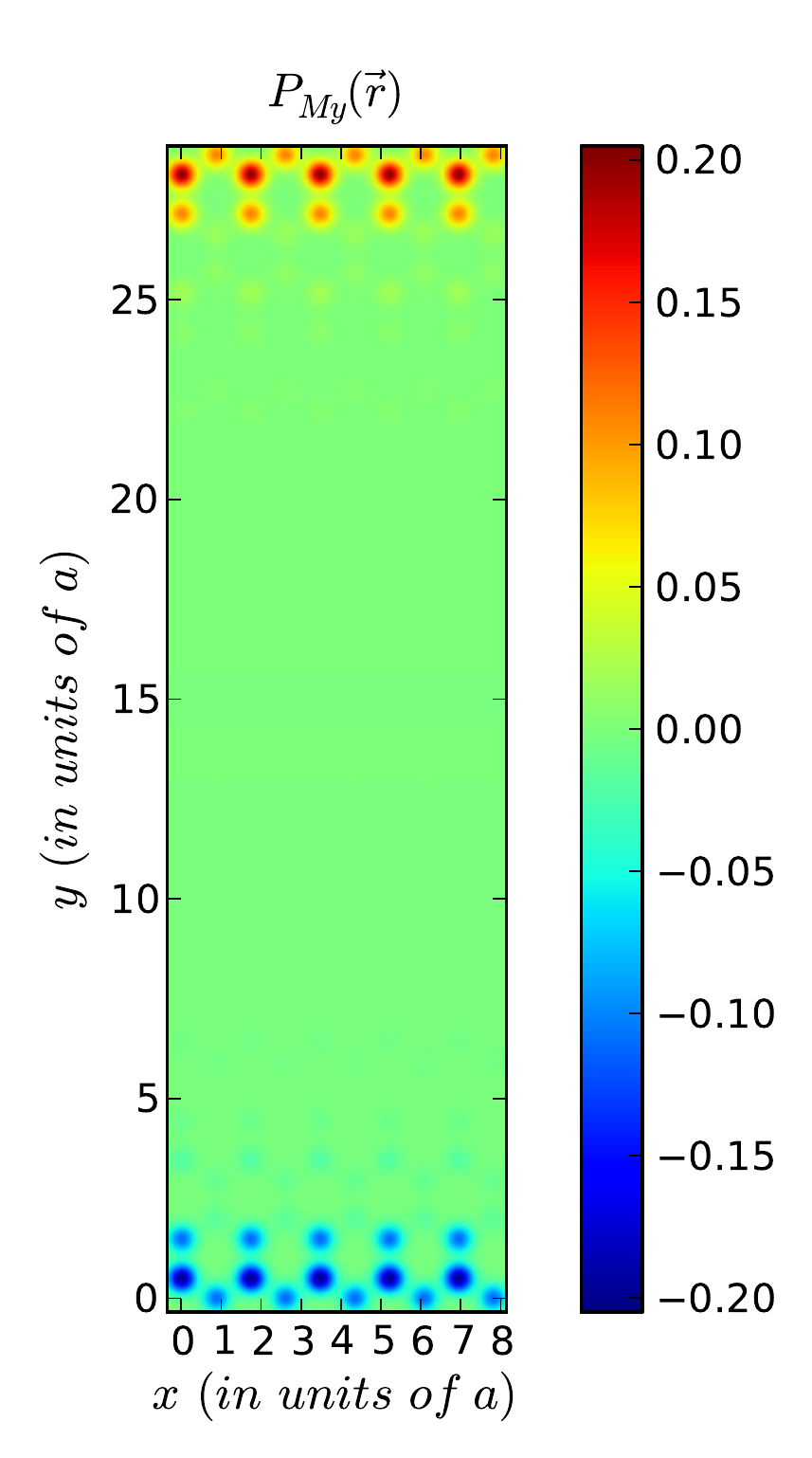}\\
\includegraphics[trim = 100mm 0mm 20mm 0mm, clip, width=4cm]{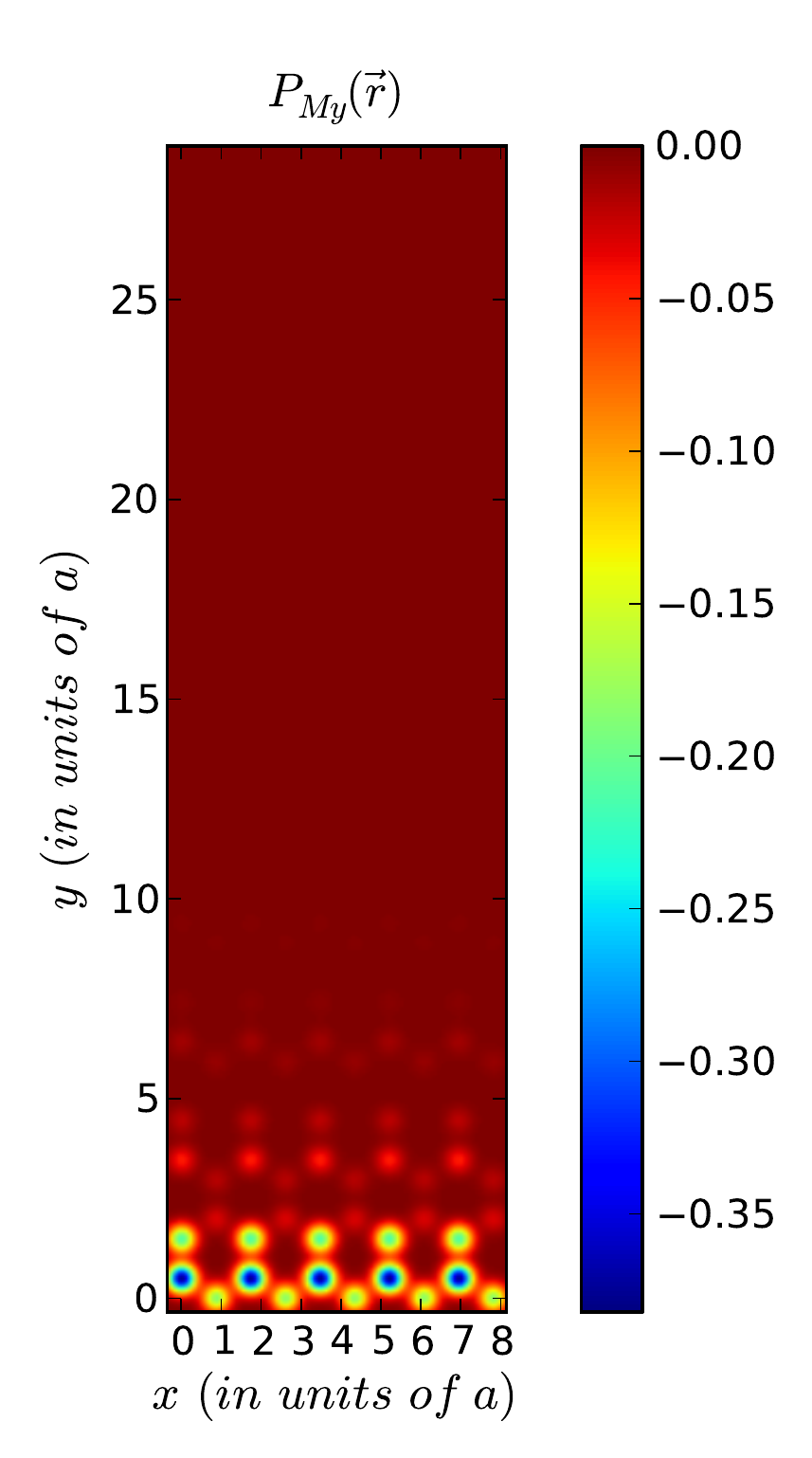}
\includegraphics[trim = 100mm 0mm 20mm 0mm, clip, width=4cm]{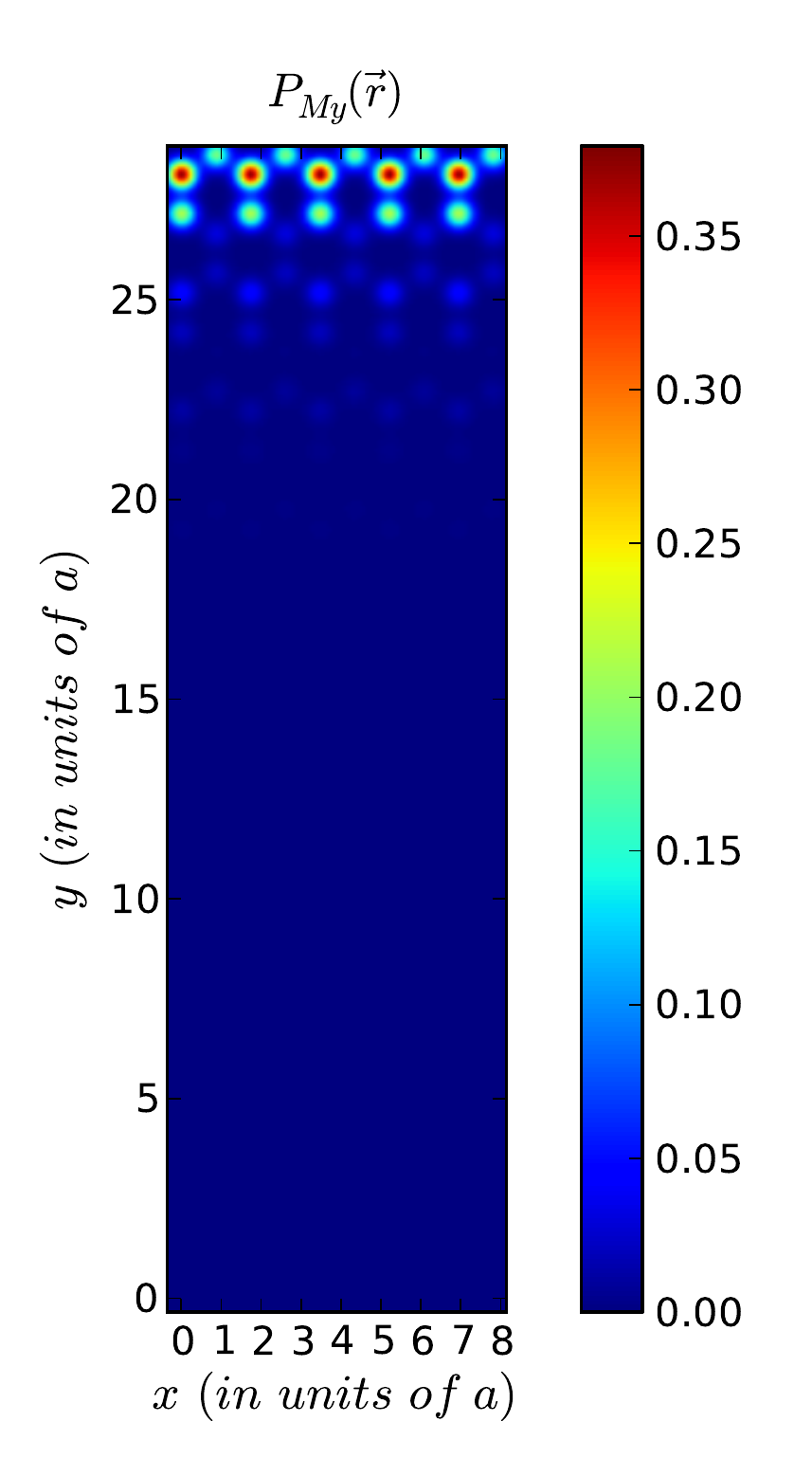}\\
\end{array}$
\caption{\small (Color online) The upper panel describes the Majorana polarization of the two zero energy modes at momentum $k_x=0$. The lower panel presents the Majorana polarization of two modes inside the gap, corresponding to a momentum of $k_x=0.3$ (resp. $k_x=-0.3$) as shown on the left panel (resp. right panel) and an energy of $E\sim0.6t$.}
\label{Majorana Polarization}
\end{figure}

The Majorana polarization $P_{M_y}$ for the zero energy states, as well as two for the highest energy states lying on the dispersive branches in the gap are presented in Fig.~\ref{Majorana Polarization}. In our chosen convention $P_{M_x}$ is equal to zero. While the $P_{M_y}$ Majorana polarization is non-zero for all of the states inside the gap, the integral of this polarization for a state centered on one of the edges gives a value of $1$ only for the zero-energy states, confirming their Majorana character. This integral is finite but not equal to $1$ for the other non-zero energy states inside the gap (this value is equal to $0.8$ for the example considered in Fig.~\ref{Majorana Polarization}) implying that they are not Majorana, but have a partial Majorana polarization similar to the Andreev bound states described in Ref.~\onlinecite{chevallier12} arising in an SNS topological junction with a finite phase difference.


While in this section we have focused mainly on the situation $t_3=2 t$ we have checked that the physics described here (i.e. the formation of the Majorana modes) is stable with respect to small deviations from this particularly symmetric point (i.e. for values of $t_3$ slightly larger or smaller than $t$) when the other parameters of the problems are kept fixed.

\subsection{Majorana modes in isotropic honeycomb-lattice structures}

\begin{figure}[t]
\centering
$\begin{array}{c}
\includegraphics[width=7cm]{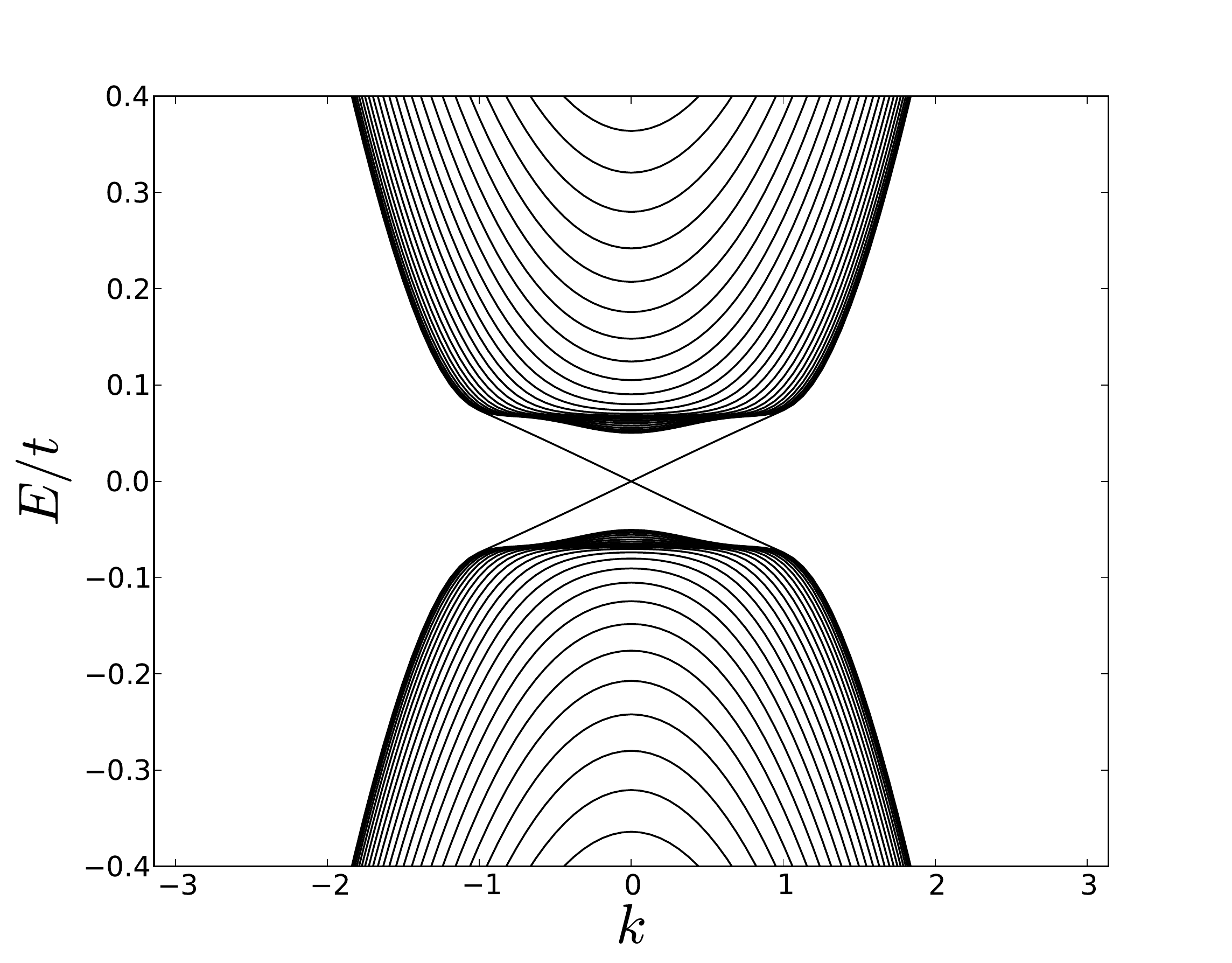}\\
\includegraphics[width=7cm]{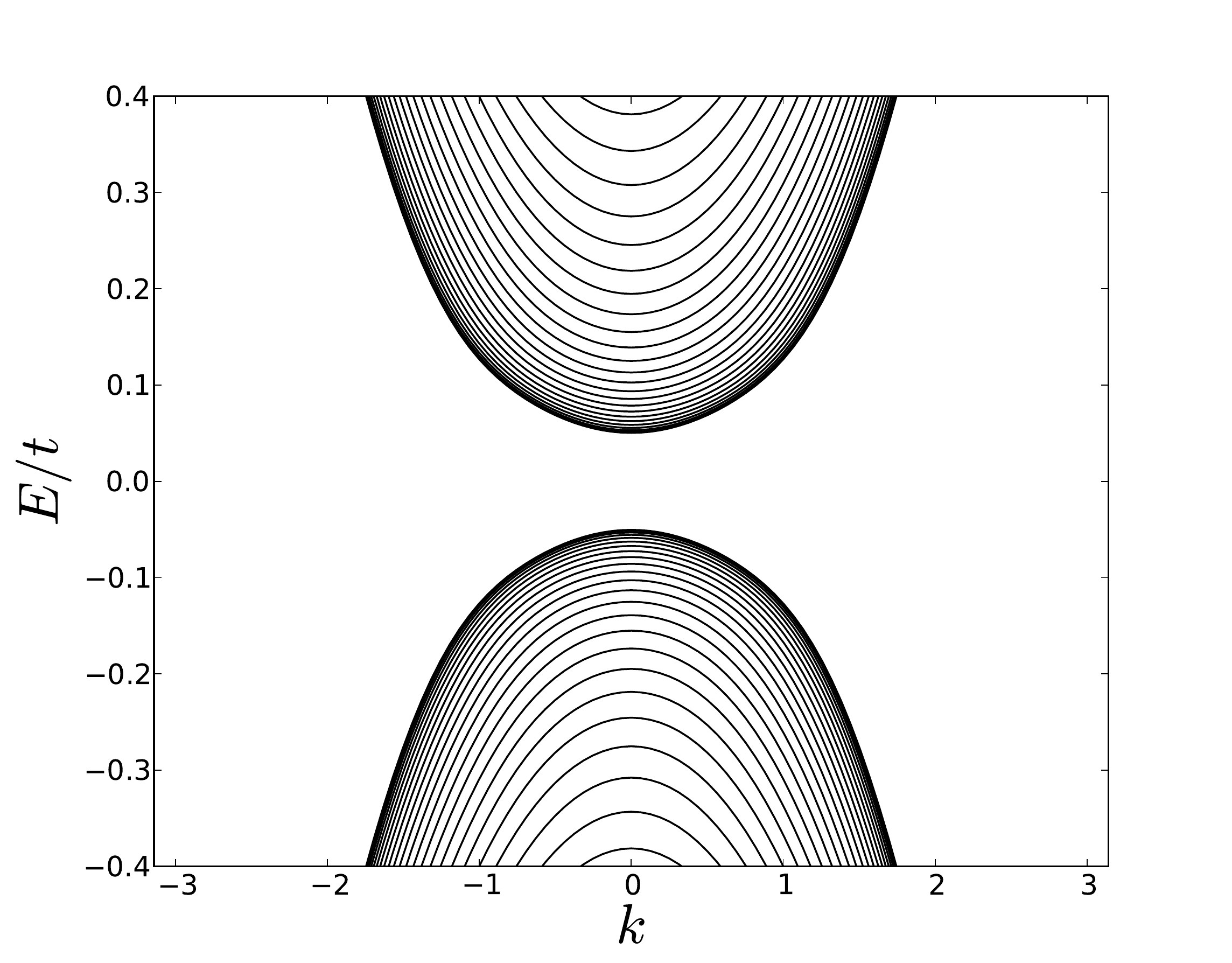}\\
\end{array}$
\caption{\small Band structure of an isotropic zigzag NR with $N=50$,  as a function of $k_{x}$. In the upper panel, the band structure in the topological phase ($\Delta=0.45t<V_Z=0.50t$), and in the lower panel the band structure in the trivial phase ($\Delta=0.55t>V_Z=0.50t$).}
\label{BandStructuresisotropic}
\end{figure}

The approach is now extended to the case of an isotropic NR, \textit{i.e.} $t_3=t$, for various values of the chemical potential. For instance, we focus on $\mu=3t$ (similarly to the anisotropic case, this corresponds to a value of chemical potential equal to the bandwidth). In this situation a similar band inversion as in the anisotropic case takes place for $\Delta < V_Z$  which leads to the formation of Majorana states inside the gap (see Fig.~\ref{BandStructuresisotropic}). While we have no analytic criterion to show that these states are indeed Majoranas for the isotropic situation, a numerical analysis of their Majorana polarization as well as the structure of their wave functions shows that this is indeed the case.

\begin{figure}[t]
\centering
$\begin{array}{c}
\includegraphics[width=7cm]{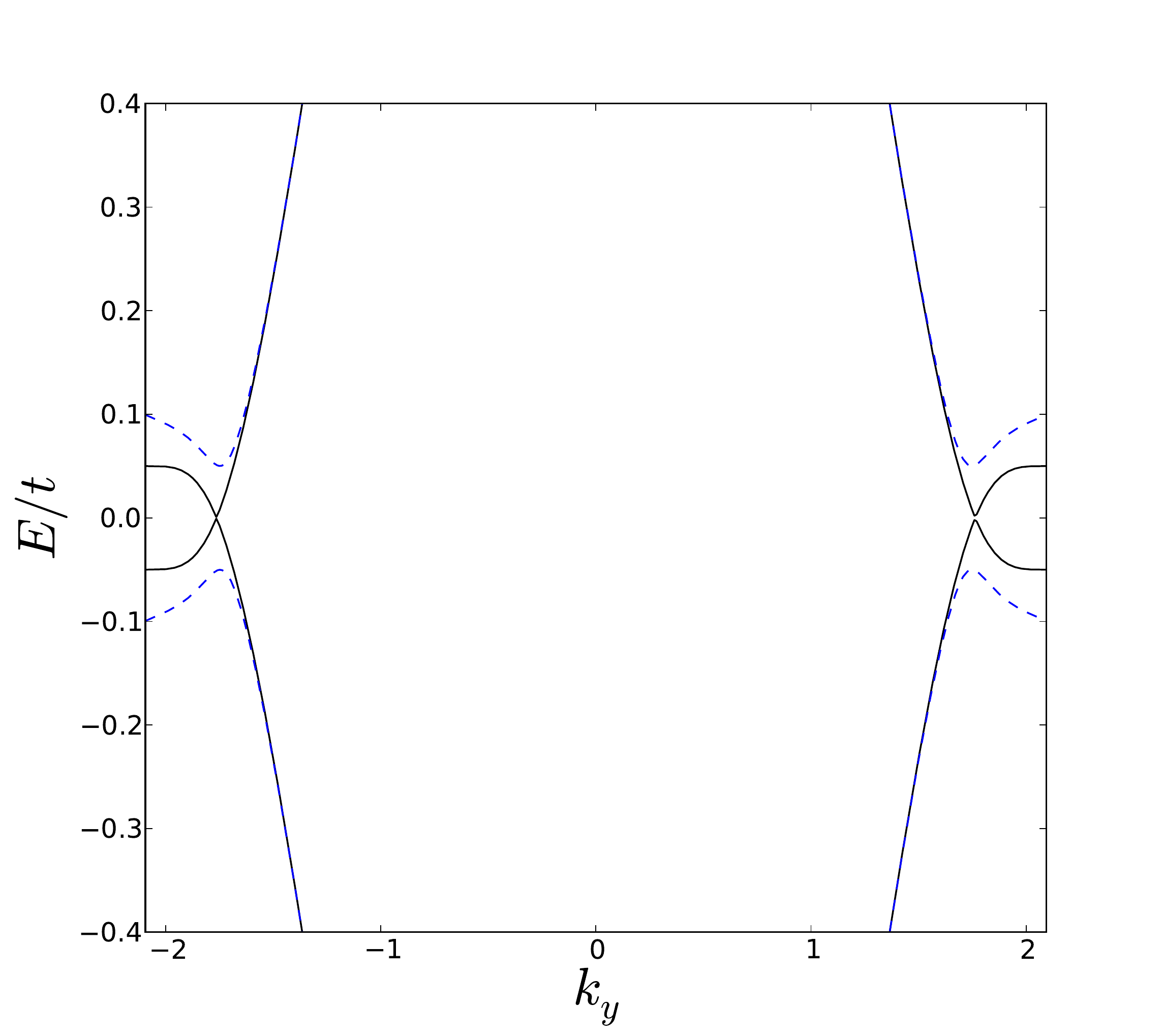}\\
\includegraphics[width=7cm]{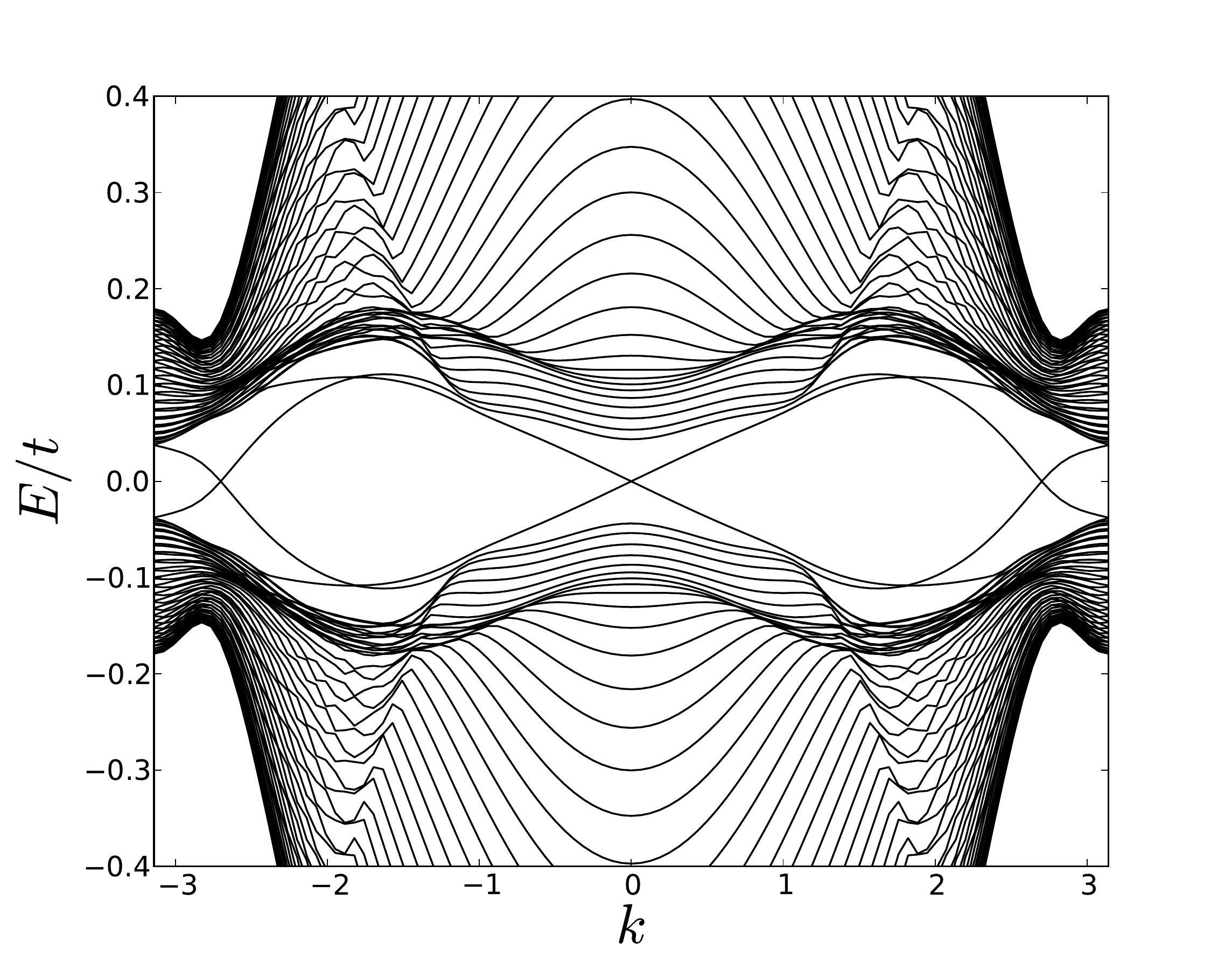}\\
\end{array}$
\caption{\small (Color online) Band structure of an isotropic infinite honeycomb lattice (upper panel) and of an isotropic zigzag NR (lower panel). The system is in the topological phase with $\lambda=0.1 t$, $\Delta=0.4 t$, $V_Z=0.5 t$ and $\mu=t$.}
\label{BandStructuresisotropicbis}
\end{figure}

Moreover Majorana fermions can arise for different values of the chemical potentials, some of them smaller and closer to more realistic values. For example, zero energy Majorana states form at $\mu=t$ (close to the Van Hove singularities), as depicted in the right panel of Fig.~\ref{BandStructuresisotropicbis}. While not shown here, Majorana dispersive states form for the same parameter values also if the ribbon has armchair edges.


\subsection{Stability of Majorana modes in presence of disorder}

\begin{figure}[t]
\centering
$\begin{array}{c}
\includegraphics[width=4.5cm]{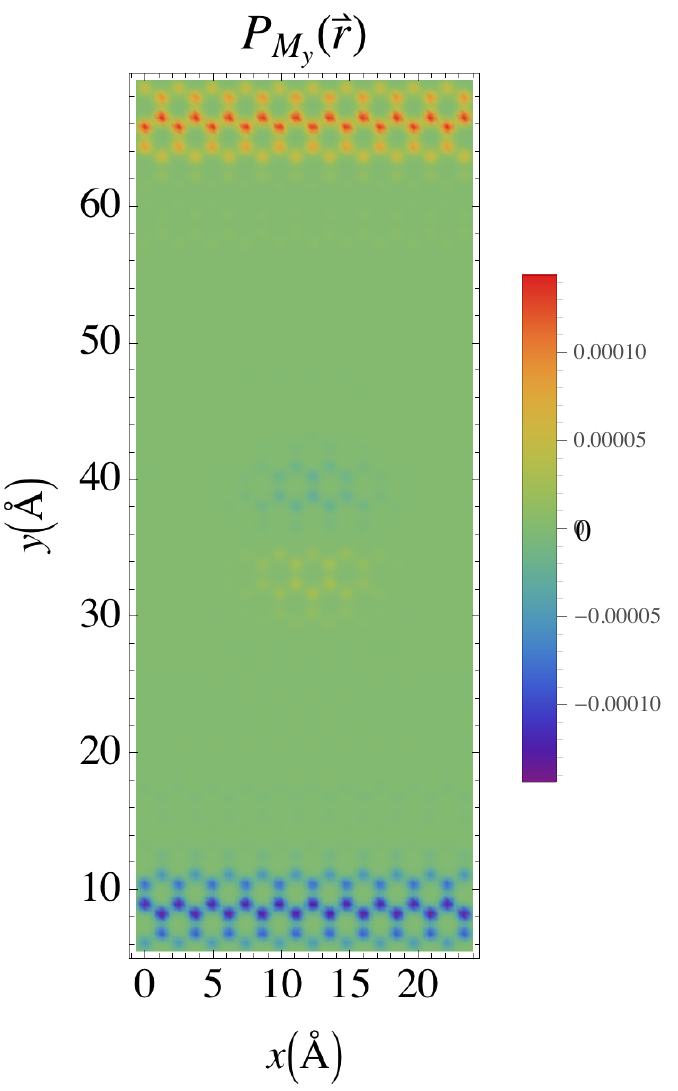}
\includegraphics[width=4.5cm]{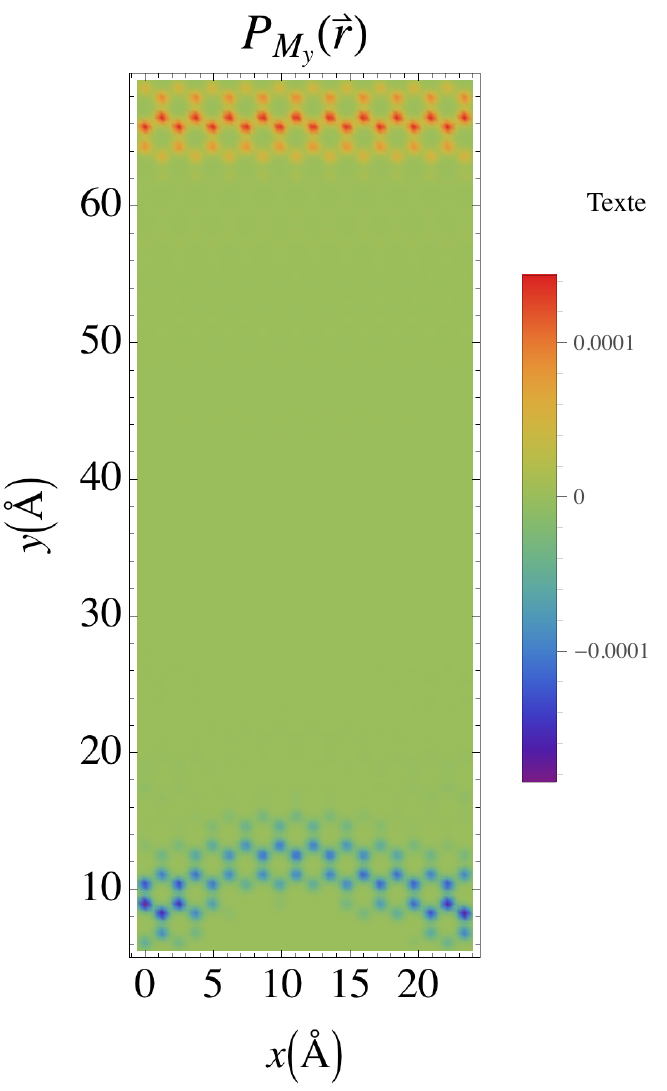}
\end{array}$
\caption{\small (Color online) Zero-energy Majorana polarization in the presence of an impurity (left panel) and in the presence of irregular edges (right panel). The system is in the topological phase with $\lambda=0.1 t$, $\Delta=0.45 t$, $V_Z=0.5 t$ and $\mu=3t$.}
\label{MajoranaPolarizationDisorder}
\end{figure}

In this section, we test the stability of the Majorana modes in presence of two types of disorder. First, we consider a single impurity in the nanoribbon, we model the impurity as an on-site shift of the chemical potential. The impurity acts as an island on which the topological condition is violated. As shown in the left panel of Fig. \ref{MajoranaPolarizationDisorder}, a localized impurity state forms in the vicinity of the impurity, exhibiting a dipole Majorana-polarization structure. A similar structure has been observed in the presence of disorder in one-dimensional wires \cite{sticlet12}.
However, we show that the edge structure is not affected by this type of bulk disorder. We have checked that the integral of the Majorana polarization is conserved and equal to one in this situation, same as in the absence of the impurity. Thus we expect that the Majorana modes that we found are stable with respect to bulk disorder.

Then, we consider a ribbon with irregular edges, as presented in the right panel of Fig. \ref{MajoranaPolarizationDisorder}. Same as in the previous case, the irregularity is modeled with the help of a fluctuating chemical potential. We find that the Majorana modes are stable also with respect to this kind of disorder. The Majorana polarization shows how the formation of the Majorana mode along the edges of the ribbon is independent of the local armchair or zig-zag nature of the edge. We have checked that the integral of the Majorana polarization is indeed conserved in this situation. 
\section{Conclusion}
We have studied the possibility to form Majorana edge states in isotropic and anisotropic nanoribbons in the presence of Rashba spin-orbit coupling, a Zeeman field and in the proximity of an s-wave superconductor. We have described a mapping between a zigzag NR and a one-dimensional dimer chain. This reduces to a 1D chain of monomers for a particular value of anisotropy for which the two Dirac cones merge into a single one. In this limit we can thus describe exactly the conditions to obtain Majorana fermionic states. 
We have tested  numerically the formation of Majorana states by using a tight-binding model and an exact diagonalization technique, as well as the Majorana polarization. We have provided a few examples for the formation of Majorana fermionic states in both isotropic and anisotropic honeycomb lattice structures with zigzag and armchair edges.

Our analysis shows that for Rashba spin-orbit couplings of the order of $0.1 t$ and for a doping corresponding to $\mu=t$, Majorana fermions can form in isotropic nanoribbons. It remains to be seen if, and for which systems, large enough dopings and spin-orbit couplings can be achieved experimentally to attain the regime of formation for the Majorana fermions. Achieving the corresponding spin-orbit coupling value in graphene for example does not seem to be an insurmountable barrier. We have shown that any small but finite value for the Rashba spin-orbit coupling would suffice to form Majorana states; moreover it has been proposed that Rashba couplings of the order of $0.2$eV can be induced in graphene using adatoms\cite{Hu12}. On the other hand it would be hard to dope graphene to chemical potentials of order $t$, such dopings may be eventually achievable in the future in other systems such as cold-atom hexagonal-lattice structures. Our work also opens the perspective to look for other spin-orbit or magnetic mechanisms that can also be envisaged to give rise to Majorana states in two-dimensional honeycomb-lattice structures. 

\section*{Acknowledgments}

The authors would like to thank P. Simon and F. Pi\'echon for stimulating discussions. The work of C.B., M.G., and D.C. is supported by the ERC Starting Independent Researcher Grant NANOGRAPHENE 256965.

\bibliographystyle{apsrev4-1}

\begin{thebibliography}{99}

\bibitem{ettore} E. Majorana, Il Nuovo Cimento (1924-1942) 14, 171 (1937).

\bibitem{beenakker} C. W. J. Beenakker,  Annu. Rev. Con. Mat. Phys. {\bf 4}, 113 (2013).

\bibitem{alicea} J. Alicea, Rep. Prog. Phys. {\bf 75}, 076501 (2012).
\bibitem{nayak_dassarma} C. Nayak, S. H. Simon, A. Stern, M. Freedman, and S. Das Sarma, Rev. Mod. Phys. {\bf 80}, 1083 (2008).
\bibitem{Moore91} G. Moore and N. Read, Nucl. Phys. {\bf B360}, 362 (1991)
\bibitem{Read00} N. Read and D. Green, Phys. Rev. B {\bf 61}, 10267 (2000)
\bibitem{Fu08} L. Fu and C. L. Kane, Phys. Rev. Lett.{\bf 100}, 096407 (2008).
\bibitem{Tanaka09} Y. Tanaka, T. Yokoyama and N. Nagaosa, Phys. Rev. Lett. {\bf 103}, 107002 (2009).
\bibitem{Sato09} M. Sato and S. Fujimoto, Phys. Rev. B {\bf 79}, 094504 (2009).
\bibitem{Potter11} A. C. Potter and P. A. Lee, Phys. Rev. B {\bf 83}, 094525 (2011).
\bibitem{Wong12} C. L. M. Wong, J. Liu, K. T. Law and P. A. Lee, arXiv:1206.5601 (2012).
\bibitem{Cook11} A. Cook and M. Franz, Phys. Rev. B {\bf 84}, 201105(R) (2011).
\bibitem{lutchyn_dassarma} R. M. Lutchyn, J. D. Sau, and S. Das Sarma, Phys. Rev. Lett. {\bf 105}, 077001 (2010).

\bibitem{oreg_vonoppen} Y. Oreg, G. Refael, and F. von Oppen, Phys. Rev. Lett. {\bf 105}, 177002 (2010).

\bibitem{mourik12} V. Mourik, K. Zuo, S. M. Frolov, S. R. Plissard, E. P. A. M. Bakkers and L. P. Kouwenhoven, Science {\bf 336}, 1003 (2012).

\bibitem{kitaev} A. Y. Kitaev, Physics-Uspekhi {\bf 44}, 131 (2001).

\bibitem{2D_diffusive} K. Bjornson and A. M. Black-Schaffer, Phys. Rev. B {\bf 88}, 024501 (2013).

\bibitem{mudry} C. Chamon, C.-Y. Hou, C. Mudry, S. Ryu, and L. Santos, Phys. Scr. {\bf T146} 014013 (2012).

\bibitem{Sato11} M. Sato, Y. Tanaka, K. Yada and T. Yokoyama, Phys. Rev. B {\bf 83}, 224511 (2011).
\bibitem{blackschaffer} A. M. Black-Schaffer, Phys. Rev. Lett. {\bf 109}, 197001 (2012).
\bibitem{klinovaja12} J. Klinovaja, G. J. Ferreira and D. Loss, Phys. Rev. B {\bf 86}, 235416 (2012).
\bibitem{klinovaja13a} J. Klinovaja and D. Loss, Phys. Rev. X {\bf 3}, 011008 (2013).
\bibitem{klinovaja13b} J. Klinovaja and D. Loss, Phys. Rev. B {\bf 88}, 075404 (2013).

\bibitem{novoselov_geim} K. S. Novoselov, A. K. Geim, V. S. Morozov, D. Jiang, Y. Zhang, S. V. Dubonos, I. V. Grigorieva and A. A. Firlsov, Science {\bf 306}, 666 (2004); A. K. Geim and K. S. Novoselov, Nat. Mater. {\bf 6}, 183 (2007).

\bibitem{honeycomb lattice_general_work} A. H. Castro Neto, F. Guinea, N. M. R. Peres, K. S. Novoselov and A. K. Geim, Rev. Mod. Phys. {\bf 81}, 109 (2009).
\bibitem{hasegawa06} Y. Hasegawa, R. Konno, H. Nakano and M. Kohmoto, Phys. Rev. B {\bf 74}, 033413 (2006).
\bibitem{montambaux09} G. Montambaux, F. Pi\'echon, J.-N. Fuchs and M. O. Goerbig, Phys. Rev. B {\bf 80}, 153412 (2009).
\bibitem{sticlet12} D. Sticlet, C. Bena and P. Simon, Phys. Rev. Lett. {\bf 108}, 096802 (2012).
\bibitem{chevallier12} D. Chevallier, D. Sticlet, P. Simon and C. Bena, Phys. Rev. B {\bf 85}, 235307 (2012). 
\bibitem{Huertas06} D. Huertas-Hernando, F. Guinea and A. Brataas, Phys. Rev. B {\bf 74}, 155426 (2006).
\bibitem{min06} H. Min, J. E. Hill, N. A. Sinitsyn, B. R. Sahu, L. Kleinman and A. H. MacDonald, Phys. Rev. B {\bf 74}, 165310 (2006).
\bibitem{yao07} Y. Yao, F. Ye, X.-L. Qi, S.-C. Zhang and Z. Fang, Phys. Rev. B {\bf 75}, 041401 (2007)

\bibitem{Efetov10} D. K. Efetov and Ph. Kim, Phys. Rev. Lett. {\bf 105}, 256805 (2010).
\bibitem{Zarea09} M. Zarea and N. Sandler, Phys. Rev. B {\bf 79}, 165442 (2009).

\bibitem{McChesney10} J. L. McChesney, A. Bostwick, T. Ohta, T. Seyller, K. Horn, J. Gonzalez and E. Rotenberg, Phys. Rev. Lett. {\bf 104}, 136803 (2010).
\bibitem{Shestov12} O. Shevtsov, P. Carmier, C. Groth, X. Waintal and D. Carpentier, Phys. Rev. B {\bf 85}, 245441 (2012).
\bibitem{Hu12} J. Hu, J. Alicea, R. Wu and M. Franz, Phys. Rev. Lett. {\bf 109}, 266801 (2012).


%
%
%
%
%
%
\end{thebibliography}

\end{document}